\documentclass[]{IEEEtran}

\usepackage{graphicx}
\usepackage{amssymb}
\usepackage{amsmath,amssymb,amsfonts}
\usepackage{epstopdf}
\usepackage{amsthm}
\usepackage{braket}
\usepackage[dvipsnames]{xcolor}
\usepackage{algorithm}
\usepackage{cite}
\usepackage[utf8]{inputenc}
\usepackage[english]{babel}
\usepackage{algpseudocode}
\usepackage[utf8]{inputenc}

\usepackage{float}
\usepackage{afterpage}

\usepackage{tabularx}
\usepackage{booktabs}

\usepackage[subrefformat=parens, labelformat=parens]{subfig}
\usepackage[font=small]{caption}
\graphicspath{{Figures/}}

\title{A Genetic Approach to Minimising Gate and Qubit Teleportations for Multi-Processor Quantum Circuit Distribution}

 \author{
      \IEEEauthorblockN{Oliver Crampton\IEEEauthorrefmark{1}, Panagiotis Promponas\IEEEauthorrefmark{1}\IEEEauthorrefmark{2}, Richard Chen\IEEEauthorrefmark{1}, Paul Polakos\IEEEauthorrefmark{1}},
      Leandros Tassiulas\IEEEauthorrefmark{2},
      Louis Samuel\IEEEauthorrefmark{1} \\
      \IEEEauthorblockA{\IEEEauthorrefmark{1}Cisco Systems
     }
      \IEEEauthorblockA{\IEEEauthorrefmark{2}Department of Electrical Engineering, Yale University
      \\}
  }

\begin{document}

\maketitle


\begin{abstract}
Distributed Quantum Computing (DQC) provides a means for scaling available quantum computation by interconnecting multiple quantum processor units (QPUs). A key challenge in this domain is efficiently allocating logical qubits from quantum circuits to the physical qubits within QPUs, a task known to be NP-hard. Traditional approaches, primarily focused on graph partitioning strategies, have sought to reduce the number of required Bell pairs for executing non-local CNOT operations, a form of gate teleportation. However, these methods have limitations in terms of efficiency and scalability. Addressing this, our work jointly considers gate and qubit teleportations introducing a novel meta-heuristic algorithm to minimise the network cost of executing a quantum circuit. By allowing dynamic reallocation of qubits along with gate teleportations during circuit execution, our method significantly enhances the overall efficacy and potential scalability of DQC frameworks. In our numerical analysis, we demonstrate that integrating qubit teleportations into our genetic algorithm for optimising circuit blocking reduces the required resources, specifically the number of EPR pairs, compared to traditional graph partitioning methods. Our results, derived from both benchmark and randomly generated circuits, show that as circuit complexity increases—demanding more qubit teleportations—our approach effectively optimises these teleportations throughout the execution, thereby enhancing performance through strategic circuit partitioning.
This is a step forward in the pursuit of a global quantum compiler which will ultimately enable the efficient use of a 'quantum data center' in the future. 
\end{abstract}

\section{Introduction}

Quantum computers can, in principle, perform tasks that have previously been impossible or highly inefficient on classical computers \cite{grover1996fast}, such as factoring large numbers using Shor's algorithm \cite{shor1999polynomial,skosana2021demonstration}, or simulating quantum systems\cite{brown2010using,daley2022practical}. However, the scaling of monolithic quantum processors towards doing useful, error free computations is difficult to achieve\cite{van2016path}. Therefore, companies such as IBM are looking towards inter-connected distributed quantum processor units (QPUs) and thus, quantum networking is required to execute useful circuits, at scale.


\begin{figure}[!t]
\centering
\subfloat[]{
\includegraphics[height=1.1in]{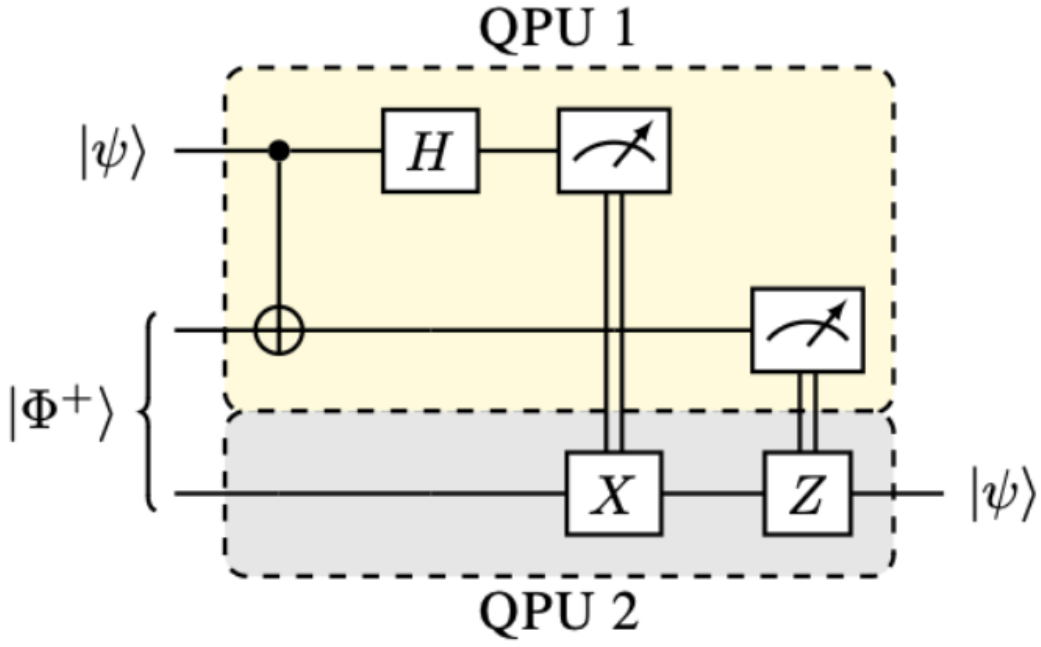}
}
\subfloat[]{
\includegraphics[height=1.2in]{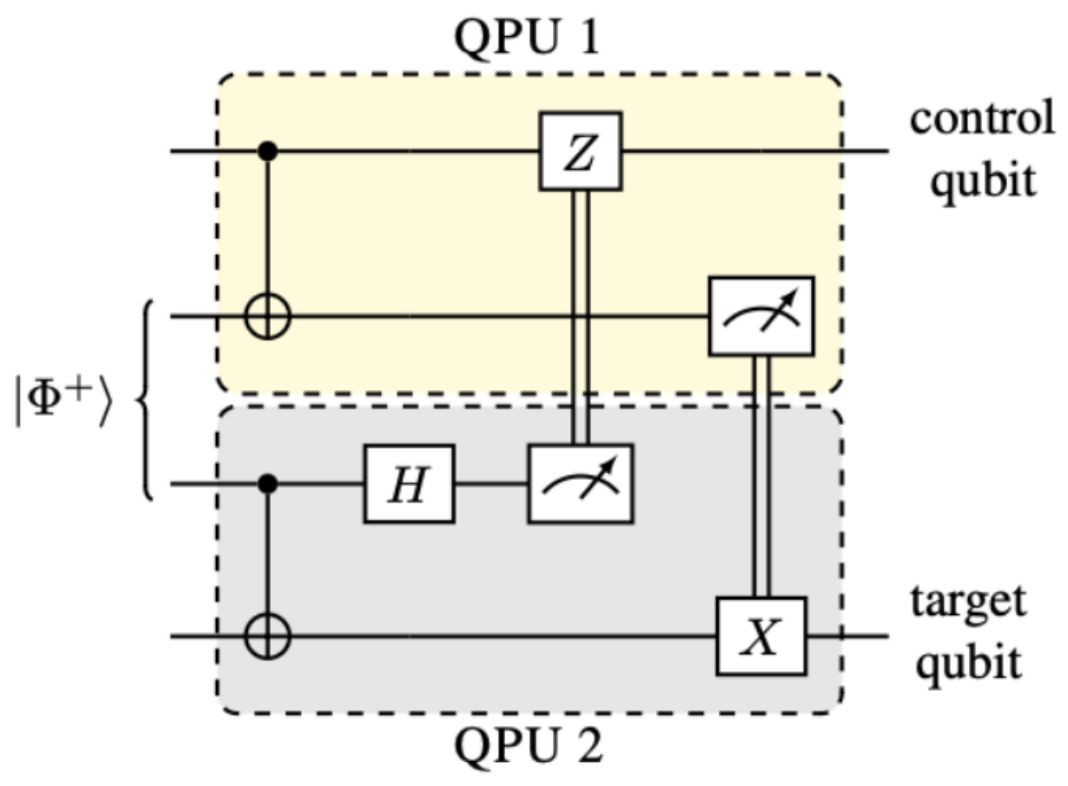}
}
\caption{Circuits that implement (a) a teleportation operation of a state $\ket{\psi}$ from QPU 1 to QPU 2, and (b) a teleportation of a CNOT operation between a \emph{control qubit} and \emph{target qubit} that are stored in different QPUs. Both operations require a Bell state $\ket{\Phi^{+}}$ as well as the transmission of classical bits that correspond to the outcome of qubit measurements.}
\label{fig:teleportations}
\end{figure}

Quantum circuits are a visual way to represent the temporal order of single or multi-qubit gates, to perform a designed algorithm. Quantum gates are unitary operations that operate on a logical qubit state $\ket{\psi} = \alpha\ket{0} + \beta \ket{1}$ \cite{nielsen2001quantum}. In Distributed Quantum Computing (DQC) there are three important types of operations, single-qubit gates (e.g., Pauli rotation, Hadamard), local CNOT (cx) (control and target qubits within the same processor), and non-local CNOT (also called telegate). In the latter, a CNOT operation should be executed between qubits that are stored in different QPUs. The Hadamard gate, Pauli gates, and CNOT operations form a universal set for quantum computation \cite{deutsch1985quantum}, meaning that any arbitrary unitary transformation of a quantum state can be expressed by only these three gates \cite{barenco1995elementary}, henceforth we will assume that all circuits have been decomposed into this universal set of operations. The CNOT gate is a two qubit operator which requires the control of one qubit by another and is given by the following matrix:

\[
CNOT = \begin{bmatrix} 1 & 0 & 0 & 0 \\ 0 & 1 & 0 & 0 \\ 0 & 0 & 0 & 1 \\ 0 & 0 & 1 & 0 \end{bmatrix}.
\]

To execute a distributed algorithm, the logical qubits must be mapped to physical qubits in the QPUs and a means for performing operations between non-local qubits must be established. Two types of non-local operations can occur, qubit or gate teleportation. Qubit teleportation is the process of transferring a logical qubit state from one QPU to another via the use of a Bell state (EPR pair) \cite{einstein1935can} and some classical communication. Gate teleportation, on the other hand, does not move the qubit state but allows one qubit to control the operation on another, distant qubit. The circuits required to perform both qubit teleportaions and a teleported controlled operation between two distant logical qubits
\cite{bennett1993teleporting}  are shown in Figure \ref{fig:teleportations}. Both operations make use of Bell states which need to be distributed by the network (one half to each QPU). This is a costly procedure and hence there is a need for minimising the teleportations in order to reduce the load on the network and to maximise the likelihood of successfully executing a quantum circuit before qubits decohere.


When executing a quantum circuit on a single QPU, it's crucial for the compiler to dynamically map logical qubits to neighboring physical positions. This mapping allows gate operations by enabling direct interactions. Numerous studies have addressed the challenges and solutions related to quantum circuit compilation (e.g., \cite{botea2018complexity,moro2021quantum,zhu2020exact}) in the case of a single QPU.  Our work, however, assumes a fully connected architecture for QPUs, where each qubit can directly interact with any other. A similar assumption would be the existence of an efficient compiler that handles the qubit mapping inside each QPU separately. Such assumptions allow us to abstract away the constraints of the compilation problem, focusing instead on optimising the network operations necessary for the distributed quantum computation. Assuming only gate teleportations as a means towards DQC, previous works have used various heuristic methods to minimise solely the number of non-local (controlled) operations within a circuit \cite{houshmand2020evolutionary,daei2020optimized, mao2023qubit, finigan2018qubit,ash2019qure}.

Although previous works have primarily considered the minimisation of gate teleporations, this work sets out to jointly consider gate and qubit teleportations as an enablement of DQC. Thus, the network cost is associated with the Bell pairs requested from both teleportation operations. Recently the authors in \cite{davis2023towards} introduced a graph partitioning framework that incorporates qubit teleportations into network cost calculations. While their work also recognizes the importance of qubit teleportations in DQC, it differs from ours in its constraint of equalizing operation counts across QPUs. Our study, in contrast, views the generation of Bell pairs for network operations as the primary limiting factor, thereby allowing for more flexible QPU operation allocations. Recently, \cite{bandic2023mapping}  employs Quadratic Unconstrained Binary Optimisation to minimise the network cost assuming only qubit teleportations. In the latter work, the authors divide a quantum circuit into predetermined slices such that each such slice can be run without the need of gate teleportations. Finally, in \cite{nikahd2021automated}, the authors employ a window based partitioning of a quantum circuit  considering also qubit teleportations. Nevertheless, the optimisation over the latter is realized through a tuning parameter that determines how ''hard" should be for a qubit to migrate to a different QPU. In contrast, in our work we optimise over the slicing of the circuit by allowing gate teleportations within each slice and qubit teleportations across slices to facilitate the distributed quantum computation.


Specifically, this paper addresses the problem of modeling and minimising the network cost of executing a quantum circuit into a DQC framework. Allowing both gate and qubit teleportations, we \emph{dynamically} allocate the logical qubits to physical qubits into the quantum processors to execute a quantum algorithm distributedly. Since both teleportation operations require a Bell pair our goal is to minimise the number of teleportations needed to complete the execution. For this purpose, we propose a novel meta-heuristic called Optimised Distributed Quantum Circuit Execution via Meta-Heuristic Approach (ODQC-MHA) that uses a genetic algorithm. Our method significantly enhances the overall efficacy and potential scalability of the DQC framework by dynamically allocating the qubits across distributed QPUs.

The rest of the paper is organised as follows, in Section \ref{sec:gate_teleportation} we introduce the logical qubit allocation problem for static assignment within a monolithic QPU as well as a straight forward partitioning heuristic. Section \ref{sec:CHUNK} introduces qubit teleportations to the qubit mapping problem and we describe our meta-heuristic for solving this problem approximately. Section \ref{sec:Performance} shows the results of the performance of our meta-heuristic against benchmark circuits and randomly generated circuits. Finally Section \ref{sec:conclusion} concludes the paper.


\section{Qubit Allocation to Minimise Gate Teleportations}
\label{sec:gate_teleportation}


Focusing only on gate teleportations as the means towards DQC, one way to decrease the network cost is to leverage graph partitioning algorithms \cite{kernighan1970efficient,fiduccia1988linear,karypis1999multilevel} in an appropriately generated graph.
In this section, we describe such process for the case of two and multiple QPUs (Section \ref{sec:gate_teleportation_KL} and \ref{sec:gate_teleportation_greedy} respectively).

In this study, we operate under the assumption that there is complete connectivity both between and within the QPUs. This means we overlook any compilation within a QPU and presume that the compiler handles the required swaps in a non-fully connected QPU to ensure adjacent qubits. Research has been done on simulating circuits on realistic, constrained processor architectures by minimising the number of swaps required to execute controlled operations \cite{childs2019circuit}. In practice constraints on the connectivity within a processor are likely to exist, however, one would need a global compiler to be able to maximise the ability to execute a circuit on NISQ devices. 

\begin{figure}
    \centering
    \includegraphics[width=0.5\textwidth,trim = 0cm 2cm 0cm 0cm, clip]{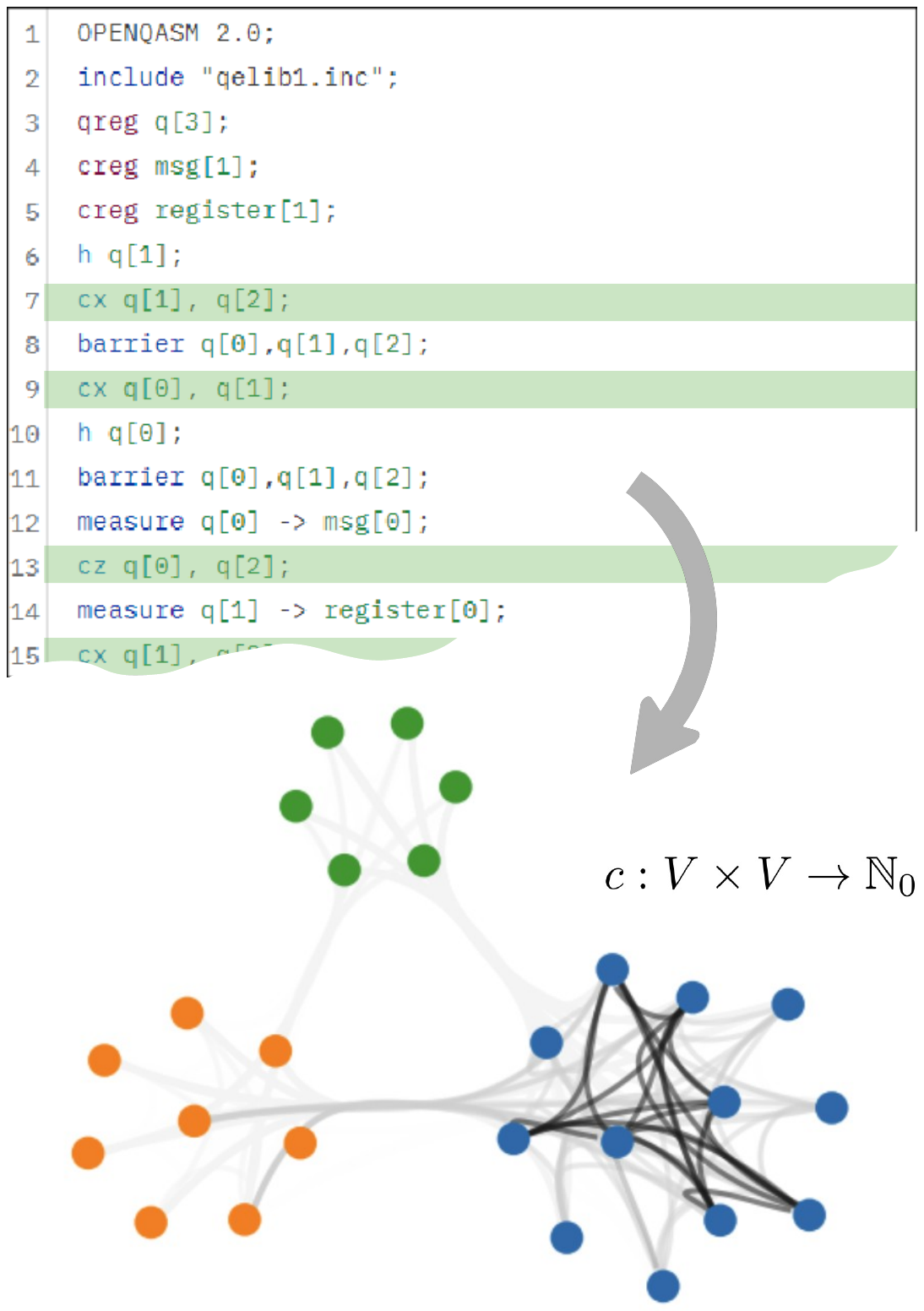}
    \caption{Example of initial qubit allocation process via graph partitioning to minimise non-local operations between 3 processors of sizes: (6,8,12). The edge weight is signified by the line colour, darker lines represent higher number of non-local operations}
\end{figure}

\subsection{Minimising Gate Teleportation: Model \& Problem Description}
\label{sec:gate_teleportation_model}



In the graph representation of a circuit denoted as $G = (V, E)$, $V$ represents a set of $n$ qubits and $E$ defines connections between qubits. The edge-weight function 
$$c: V \times V \rightarrow \mathbb{N}_0,$$ 
where $\mathbb{N}_0$ denotes the set of natural numbers including zero represents the frequency of controlled operations between qubits $u$ and $v$. Therefore, $c(u,v) = 0$ indicates the absence of an edge and thus of a CNOT gate between $u$ and $v$.
The cost of a partition $cost(V_1,V_2)$ is defined as the sum of all weights $c(u,v)$ where $u \in V_1$ and $v \in V_2$ belong to different partitions. The aim is to find $k$ partitions of the graph each of at most size $v=\frac{n}{k}$ such that the capacity of the edges between partitions is minimised, thus reducing the number non-local controlled operations, and the number of Bell pairs required. The constraint of \emph{almost} equal partitions is to minimise the maximum number of physical qubits needed from a QPU.

Minimising the number of non-local operations is crucial because entanglement is a costly resource and distributing it into a network of QPUs requires extra time steps and is error-prone \cite{abelem2023quantum}.

\subsection{K-L Algorithm for 2 QPUs}
\label{sec:gate_teleportation_KL}

One commonly used heuristic method for bi-partitioning a weighted graph is the Kernighan-Lin algorithm \cite{kernighan1970efficient}. The K-L algorithm works by taking the weighted graph $G=(V,E)$ and $c$ as the edge-weight function. By swapping pairs of vertices $u_i \in V_1$ and $v_i \in V_2$ with maximum cost improvement, the swapped pair are locked in place and the same process is done with another pair until all vertices are locked. The best configuration is chosen and the algorithm is run again until a close to optimal configuration is found. The downside of the K-L algorithm is that it can only be used for bi-partitioning of a graph. In the next section we propose a similarly simple heuristic for partitioning a graph, into any size partitions $k$. Such extension enables the division of a quantum circuit's logical qubits across multiple QPUs, beyond just two, when available.

\subsection{Greedy Partitioning Algorithm in the case of multiple QPUs}
\label{sec:gate_teleportation_greedy}

\begin{algorithm}
\caption{Greedy Partitioning Algorithm (GPA)}
\label{greedy}
\begin{algorithmic}[1]
\Procedure{AssignQubitsToQPUs}{$G, QPUList$}
    \State $Filled \leftarrow \{\}$ \Comment{Set to track filled QPUs}
    \State $Allocation \leftarrow \{\}$ \Comment{Map of qubits to QPUs}
    \State $QPUOrder \leftarrow$ sort $QPUList$ by capacity, descending
    \For{$QPU$ in $QPUOrder$}
        \State $Supernode \leftarrow$ \{\}
        \While{$Supernode$ size $<$ $QPU$ capacity and $G$ has unallocated nodes}
            \State $TargetNode \leftarrow$ select node in $G$ with max weighted edge to $Supernode$, not in $Filled$
            \State Merge $TargetNode$ into $Supernode$
            \State Update $Allocation$ to include $TargetNode \rightarrow QPU$
        \EndWhile
        \State Add $Supernode$ to $Filled$
    \EndFor
    \State \Return $Allocation$
\EndProcedure
\end{algorithmic}
\end{algorithm}

In this section we describe Greedy Partitioning Algorithm in the case of multiple QPUs (GPA),  a straightforward and practical heuristic approach for distributing qubits in circuits with varying numbers of qubits and depth, across any quantity and scale of QPUs. The heuristic works by always contracting the largest weighted edge to build supernodes - respresenting QPUs - one by one. Once a supernode (QPU) has been filled to capacity, none of the nodes within can be swapped later in the algorithm. The largest weighted edge that is adjacent to the supernode is always contracted at each step of the heuristic, with no look-ahead. This method is computationally inexpensive and so can be used as a quick heuristic for qubit allocation within our proposed meta-heuristic (proposed in Section \ref{sec:CHUNK}). The pseudo-code for GPA is shown in Algorithm \ref{greedy}. This heuristic is performed once to produce a good allocation of qubits to processors where the number of interactions between each processor is reduced. The algorithm is implemented in python and utilises the NetworkX framework to do the edge contractions. Here, an edge contraction is the process of producing a graph in which two node $v_1$ and $v_2$ are replaced with a single node, $v$, such that $v$ is adjacent to the union of the nodes to which $v_1$ and $v_2$ were originally adjacent, also called 'vertex identification'.



\section{Minimising Remote Operations: Model \& Problem Description}
\label{sec:CHUNK}



Thus far, our efforts have concentrated on reducing the quantity of non-local controlled operations, aiming ultimately to decrease the necessary number of Bell pairs throughout the execution of a distributed quantum circuit. However, if instead we aim to minimise the overall number of Bell pairs, we can allow for the reallocation of qubits within one application via qubit teleportation operations. In this section we introduce the proposed framework, Optimised Distributed Quantum Circuit Execution via Meta-Heuristic Approach (ODQC-MHA), with the goal of optimising the partition of a quantum circuit to distributed QPUs. In this section, we introduce ODQC-MHA by describing its high level design, introducing the genetic algorithm being used in the framework and finally analyzing in detail every component that it comprises (Sections \ref{sec:metaheuristic_high_level}, \ref{sec:metaheuristic_genetic_description}, \ref{sec:metaheuristic_low_level} respectively).

\subsection{Optimised Distributed Quantum Circuit Execution via Meta-Heuristic Approach (ODQC-MHA) - High Level Design}
\label{sec:metaheuristic_high_level}

In this section we introduce the ODQC-MHA framework by describing its high level design. ODQC-MHA allows the circuit to be analysed in blocks of varying size, using any algorithm for k partitioning a graph within each block. For each block we attempt to minimise the number of gate teleportations while allowing qubit teleportations between blocks to re-allocate the logical qubits when needed. Finding the optimal blocking is challenging because of the many possible combinations, however, this problem is to be solved by our heuristic. Note that each allocation has no information about the previous block's allocation and so a meta-heuristic is required to minimise the gate and qubit teleportations together. 

\begin{figure}[ht]
\centering
\includegraphics[width=0.5\textwidth,trim = 2.5cm 16cm 4cm 2.5cm, clip]{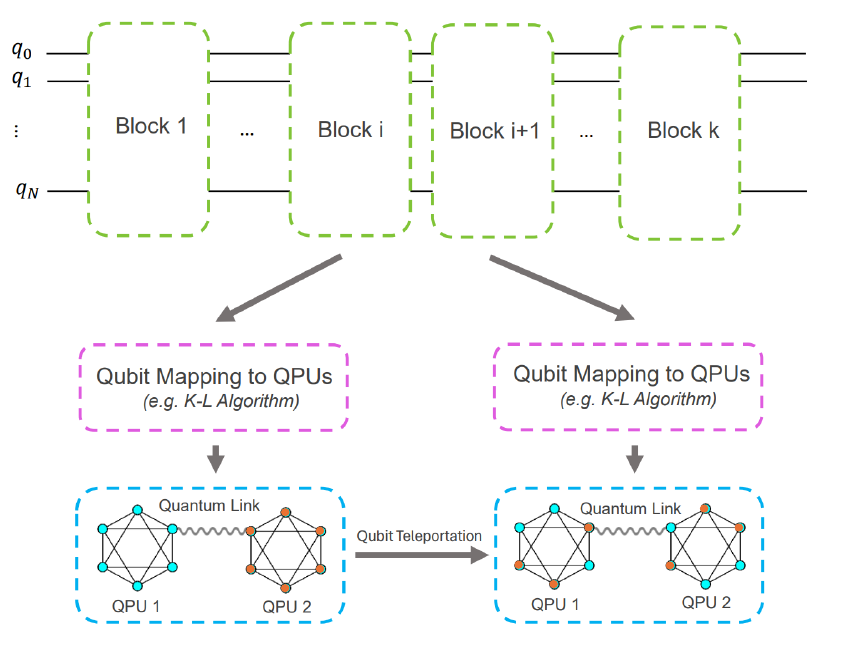}
\caption{ A high level overview diagram of the proposed framework (ODQC-MHA) for blocking/partitioning quantum circuits. The circuit is broken into blocks of arbitrary size (number of layers). Within each block, logical qubit allocation is performed using graph partitioning methods in order to reduce the number of Bell pairs required for non-local operations. After this, between each block, qubit teleportations are performed to reallocate the logical qubits according to each blocks allocation.}
\label{fig:block_algorithm}
\end{figure}

The high level description of the proposed framework is illustrated in Figure \ref{fig:block_algorithm}. Note that to enhance circuit execution efficiency, qubit teleportations are allowed between blocks. Given the vast search space comprising various circuit partitions and qubit placements, the approach combines any partitioning algorithm for intra-block qubit placements (this can be K-L, GPA or any other procedure) with a genetic algorithm to jointly consider the qubit teleportations. The proposed genetic algorithm evaluates its utility given a circuit partition based on the placements suggested by the particular partitioning algorithm used, focusing on achieving the optimisation objective of minimising the network cost. Note that this is not a joint optimisation but a meta-heuristic that uses the output of some heuristic (K-L, Greedy algorithm etc.) to explore the solution space more thoroughly. The problem of graph partitioning is NP-HARD per block hence the blocking model proposed in this paper is hard to solve without a novel heuristic.

In the next sections, a genetic algorithm is proposed to approximate an optimal 'blocking' of the circuit to minimise the total number of Bell pairs required for qubit and gate teleportations.

\subsection{Optimised Distributed Quantum Circuit Execution via Meta-Heuristic Approach (ODQC-MHA) - Genetic Algorithm}
\label{sec:metaheuristic_genetic_description}

Genetic algorithms mimic natural evolution by evolving solutions to problems through a process of selection, mutation, and crossover \cite{katoch2021review}. They start with a diverse population of individuals, where each individual's "genotype" encodes a potential solution, and its "phenotype" — its performance or fitness — reflects the solution's effectiveness. Over successive generations, individuals with higher fitness are more likely to pass their genes to the next generation, allowing the algorithm to "naturally select" increasingly effective solutions.


In our approach, we utilize a genetic algorithm to optimise the distribution of computational tasks in a quantum computing network, specifically aiming to minimise the requisite number of Bell pairs for efficient quantum communication. The core of our algorithm is defined by a population of candidate solutions, denoted as  $P = \{p_1,p_2,...,p_N\}$, where each candidate solution $p_i$ represents a potential configuration of dividing the target quantum circuit into distinct blocks.

Each candidate solution $p \in P$ is characterized by its genotype, $G_p$, which in our model is a sequence of integers  $G_p = \{g_1,g_2,...,g_K\}$, where $g_1, \dots, g_k \in \mathbb{N}$ and $K$ represents a predefined maximum number of blocks for the quantum circuit. Here, $g_i$ signifies the depth (i.e., the number of layers) of the circuit within block $i$ of the network. Note that large value for $K$ increases the search space exponentially allowing for more combinations of circuit blockings to be checked by the algorithm. Intuitively, the configuration of these blocks is subject to a constraint where the sum of all $g_i$ values must equal the total number of layers in the quantum circuit. Notably, it is permissible for any $g_i$ to be zero, indicating blocks that are empty and thus not contributing to the overall division of the circuit. For instance, a genotype $G_{p} = [12,42,64,38,203,0,34]$ represents a circuit partitioned into blocks with respective depths of $12, 42, 64, 38, 203, 34$. Although in this specific example we allow up to $7$ blocks for the circuit, this specific genotype, $G_{p}$, utilizes only $6$ of them.

The phenotype, $X_p$, associated with an individual $p \in P$ with genotype $G_p$, quantifies the total number of Bell pairs required for the candidate solution's block configuration. This total encapsulates both gate teleportations within individual blocks and qubit teleportations across the network. The phenotype thus serves as a measure of the solution's effectiveness in optimising quantum communication. To evaluate the phenotype and thus the viability of each candidate solution, we introduce a fitness function,  $evaluateFitness : P \rightarrow \mathbb{N}$, that maps the individual to a natural number. In our case, this function is counting the total number of Bell pairs and hence teleportations are needed under the configuration under consideration. 

Through the iterative processes of selection, crossover, and mutation, our genetic algorithm seeks to evolve the population towards configurations that minimise Bell pair usage, thereby enhancing the efficiency and feasibility of DQC tasks. By continuously refining the genotypes within the population based on their fitness scores, the algorithm drives towards an optimal or near-optimal distribution of computational loads and quantum communication requirements across the network. This framework not only provides a method for optimising quantum network configurations but also offers insights into the trade-offs between computational depth and quantum communication resources, laying the groundwork for further innovations in DQC architectures.

\begin{figure*}[t!]
    \centering
    \includegraphics[width=\linewidth]{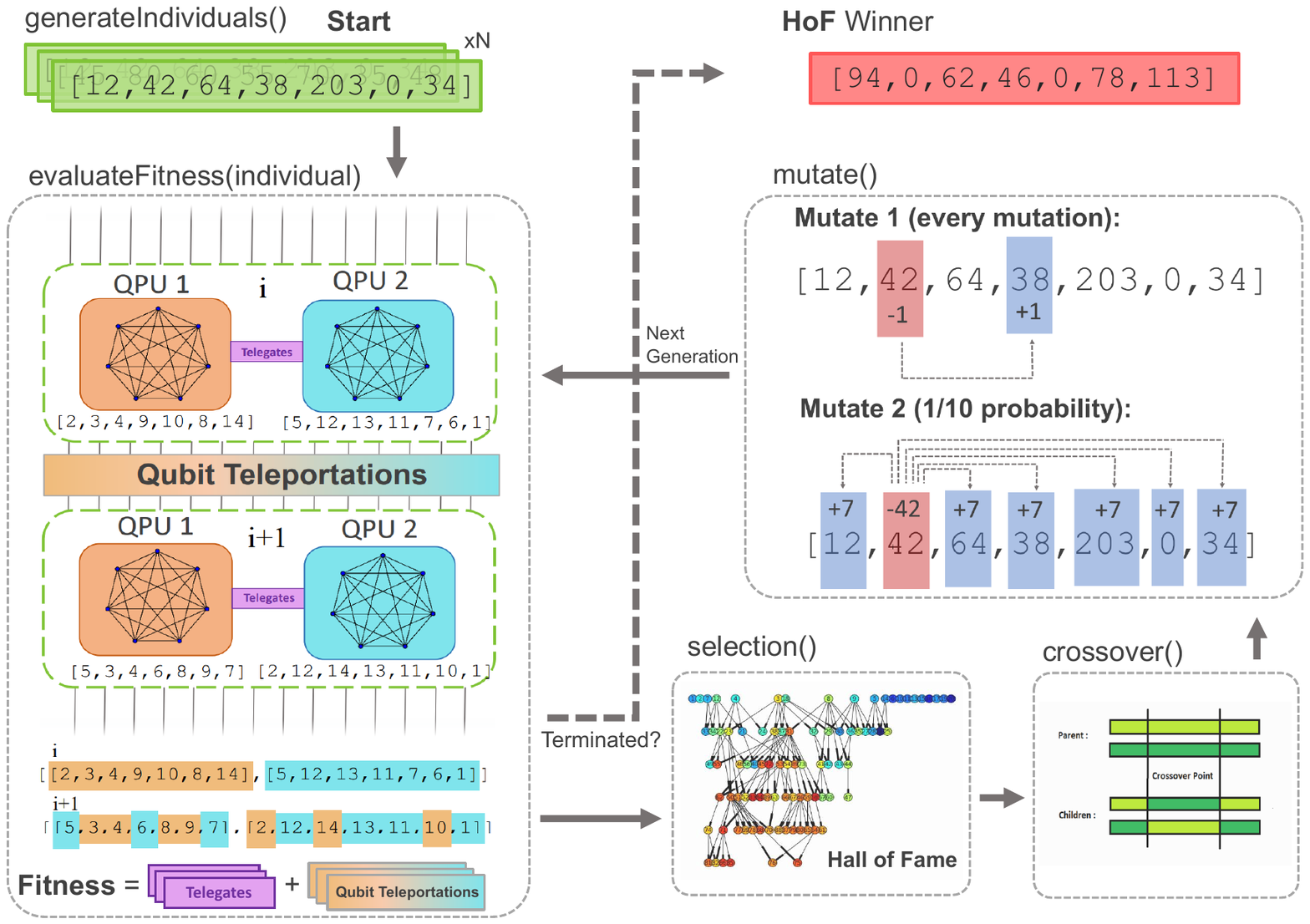}
    \caption{Structure of genetic algorithm. \emph{generateIndividuals()}: create N lists, of a given size, representing the number of layers per block of a circuit. \emph{selection()}: 'Hall of Fame' selection process ensures that the best individual to ever exist is chosen as the optimal. \emph{crossover()}: randomly chooses two individuals from the mating pool to create each new generation of superior individuals. \emph{mutate}: individuals are randomly chosen to mutate, mutate 1 occurs every mutation and mutate 2 occurs with probability 1/10 for each mutation. \emph{HoF Winner}: the best individual that has existed throughout the generations is selected as the optimal 'blocking' solution.}
    \label{fig:genetic_alg}
\end{figure*}

\subsection{Optimised Distributed Quantum Circuit Execution via Meta-Heuristic Approach (ODQC-MHA) - A Detailed Description}
\label{sec:metaheuristic_low_level}

The Optimised Distributed Quantum Circuit Execution via Meta-Heuristic Approach (ODQC-MHA) makes use of a genetic algorithm formalism to find a minimum number of total Bell pairs required for a distributed execution of a given quantum circuit by finding an arrangement of block lengths that minimises the total cost. This section provides a detailed description of the ODQC-MHA components that were abstracted away in Figure \ref{fig:block_algorithm}, complementing also the overview provided in Figure \ref{fig:genetic_alg}.

\subsubsection{Efficient Qubit Teleportation Decisions}
\begin{figure}
\centering
\includegraphics[width=0.48\textwidth,trim = 3cm 22cm 9cm 2.5cm, clip]{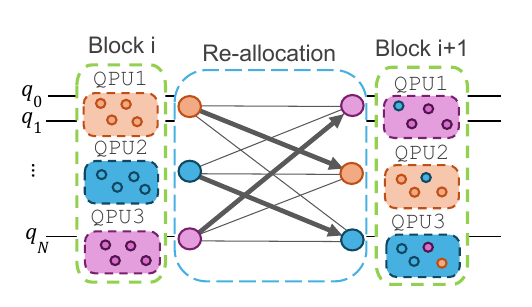}
\caption{Bipartite graph representing the re allocation between multiple QPUs, it is clear to see that there are multiple ways to map a given allocation to QPU.}
\label{fig:buipartite}
\end{figure}

For each individual in the genetic algorithm, the quantum circuit is segmented into blocks based on the number of layers specified by the individual's genotype. To optimise the allocation of qubits across multiple quantum processors (QPUs), a method such as graph partitioning (e.g., GPA) is employed. This step aims to find an approximately optimal distribution of qubits over the available processors \emph{for each block}.  Given the allocation for each block, the challenge arises in transitioning qubits from their configuration in one block to the next. This transition is not straightforward due to the flexibility in QPU assignments: any given allocation might correspond to any QPU. This leads to a complex problem, especially as the number of QPUs increases, where finding the most efficient mapping between allocations in consecutive blocks becomes computationally intensive.  To address this, we construct a bipartite graph $G=(I,J,L)$, where nodes in disjoint set $I$ represent the processor allocations in block $i$, and nodes in set $J$ represent the allocations in block $i + 1$ (Figure \ref{fig:buipartite}). The edges in this graph, $L$, denote the potential mappings between allocations, with weights reflecting the minimum number of qubit differences (and thus qubit teleportations) required for each mapping. By negating these weights and applying a maximum weighted matching algorithm \cite{schrijver2003combinatorial}, we identify the mapping that minimizes the number of qubit teleportations needed for reallocation between blocks. This approach significantly reduces the complexity of finding optimal qubit transitions between blocks.

\subsubsection{Components of the Genetic Algorithm}
\hfill
\par
\textbf{Initialisation (generateIndividuals)} - firstly the initial population is generated. Towards that goal we generate homogeneous lists of a given length, which corresponds to the maximum number of blocks available in the solution. Then applying the mutate function (described later) to each individual $100,000$ times we generate diverse genotypes for the initial population. It is important that these individuals are highly varied due to the size and complexity of the solution space.

\textbf{Evaluation of Fitness (evaluateFitness)} - 
the fitness function evaluates the efficiency of a given qubit allocation and teleportation scheme. It does so by summing the total number of gate teleportations within each block (determined by the initial qubit allocations) and the qubit teleportations between blocks (as optimised by the bipartite graph matching). The objective of the genetic algorithm is to minimize this sum, thereby reducing the overall quantum communication and computation overhead in the DQC framework.  The evaluation process effectively quantifies the "cost" of a particular configuration of qubit allocations and transitions, guiding the genetic algorithm toward solutions that optimise the use of quantum resources. By focusing on minimizing the combined total of gate and qubit teleportations, the algorithm seeks configurations that offer the best balance between computational efficiency and the practical constraints of DQC.  This structured approach allows for a clear understanding of how qubit teleportation and allocation decisions impact the overall efficiency of quantum computing operations, providing a solid basis for optimising DQC architectures.

\textbf{Crossover Function} - the crossover function is the process of generating offspring from the selected parent genes. These offspring are generated such that they share some elements from either parent. The crossover function used is a simple two-point crossover which chooses a subset of random size from each pair of parents and swap them. Since the genotype's values must sum up to the total number of layers in the quantum circuit, we employ a rebalancing routine to enforce that constraint after the crossover.
    
\textbf{Mutation Function} - The mutation function involves randomly choosing 2 indices $i$ and $j$, $i \neq j$ in the individual, with a predefined probabilty of occurance. Defining a mutation constant, c, a fair coin is flipped and the mutation constant is either added to $g_i$ and subtracted from $g_j$ or vice versa. Also each mutation has a probability $p$ of introducing a zero element to the individual, by subtracting a randomly chosen indices value from itself and spreading the value among the remaining indices.

\textbf{Selection Process} - The selection process entirely replaces the parental population, requiring that the selection procedure is stochastic and allows the same individual to be selected more than once. We used the selTournament() function as part of the DEAP framework, which was found to be a suitable mutation function for searching the solution space thoroughly.

\subsubsection{Overview}

The genetic algorithm adjusts the size of each block - between which qubit teleportations are done to get from one allocation to the next - to try and minimise the total number of Bell pairs required. The size of a block is allowed to go to zero and if so it is ignored by the calculation of teleportations. In other words, the optimal 'blocking' of the circuit may contain fewer blocks than the initial candidate solution. Note that this meta-heuristic can use any algorithm for the allocation of qubits inside a given block to calculate the Bell pairs needed for the gate teleportations. For example, if there are just two QPUs one could implement K-L algorithm and in a multi-QPU framework the proposed GPA is more suitable.

In the next section, we implement ODQC-MHA using the DEAP python framework \cite{de2012deap} to execute the genetic algorithm with these definitions, to converge on a close to optimal blocking of the circuit that required the minimum number of gate and qubit teleportations combined for a given quantum circuit. Hereafter, the term ODQC-MHA(K-L) refers to the meta-heuristic that applies the K-L algorithm to each block. Conversely, ODQC-MHA(GPA) denotes the variant where the GPA is employed for graph partitioning in each block.

\section{Performance Evaluation}
\label{sec:Performance}


As we have discussed, the objective of our heuristic is to minimise the total number of bell pairs required for distributed execution, which we believe is the bottleneck in a quantum network. To evaluate the performance we compare the number of bell pairs required for various circuits using ODQC-MHA against an existing method of qubit allocation, namely K-L.

\subsection{Pre-processing}
In this section we describe the steps taken to analyse benchmark circuits in order to evaluate the performance of ODQC-MHA. Each circuit that is analysed is represented in a QASM (QUantum ASeMbley) \cite{cross2017open} file that contains all of the logical instruction information in order. The QASM file is parsed and the circuit is then represented as a directed acyclic graph (DAG) which shows the dependencies of each gate and allows us to determine which operations can occur simultaneously (in the same layer). Given that we have a circuit in layers, we can split the circuit, by layer, into the blocks given by an individual's genotype $G_p$, as explained previously. For each block, an interaction matrix is constructed using the QASM instructions, this matrix is then used to build a networkx \cite{hagberg2008exploring} graph object for which the standard graph partitioning algorithms can be used. ODQC-MHA can be applied to any circuit of any size, although the execution time scales with the number of qubits and the maximum number of blocks. Note that the bottleneck here is the number of qubits in the circuits as this determines the size of the graph to be partitioned.

\begin{table}[ht]
\centering
\noindent \begin{tabularx}{0.5\textwidth}{l *{5}{>{\raggedleft\arraybackslash}X}}
\toprule
ID & Circuit name & Qubits & Depth (CX only) & Unary gates & CX gates \\
\midrule
1 & adder\_n118 & 118 & 4 & 1107 & 845 \\
2 & sym9\_146 & 12 & 91 & 180 & 148 \\
3 & cycle10\_2\_110 & 12 & 3386 & 3402 & 2648 \\
4 & inc\_237 & 16 & 3463 & 5983 & 4636 \\
5 & cm85a\_209 & 14 & 3818 & 6428 & 4986 \\
6 & rd84\_253 & 12 & 4466 & 7698 & 5960 \\
7 & root\_255 & 13 & 5354 & 9666 & 7493 \\
8 & mlp4\_245 & 16 & 6190 & 10620 & 8232 \\
9 & clip\_206 & 14 & 10734 & 19055 & 14772 \\
10 & dist\_223 & 13 & 11911 & 21422 & 16624 \\
\bottomrule
\end{tabularx}
\caption{Benchmark quantum circuits for Section~\ref{ssec:num_non_random_qcircuits}.}
\label{table:circuits}
\end{table}

\subsection{Non-random Quantum Circuits (2 QPUs)}
\label{ssec:num_non_random_qcircuits}
\begin{figure}
\centering
\includegraphics[width=0.5\textwidth]{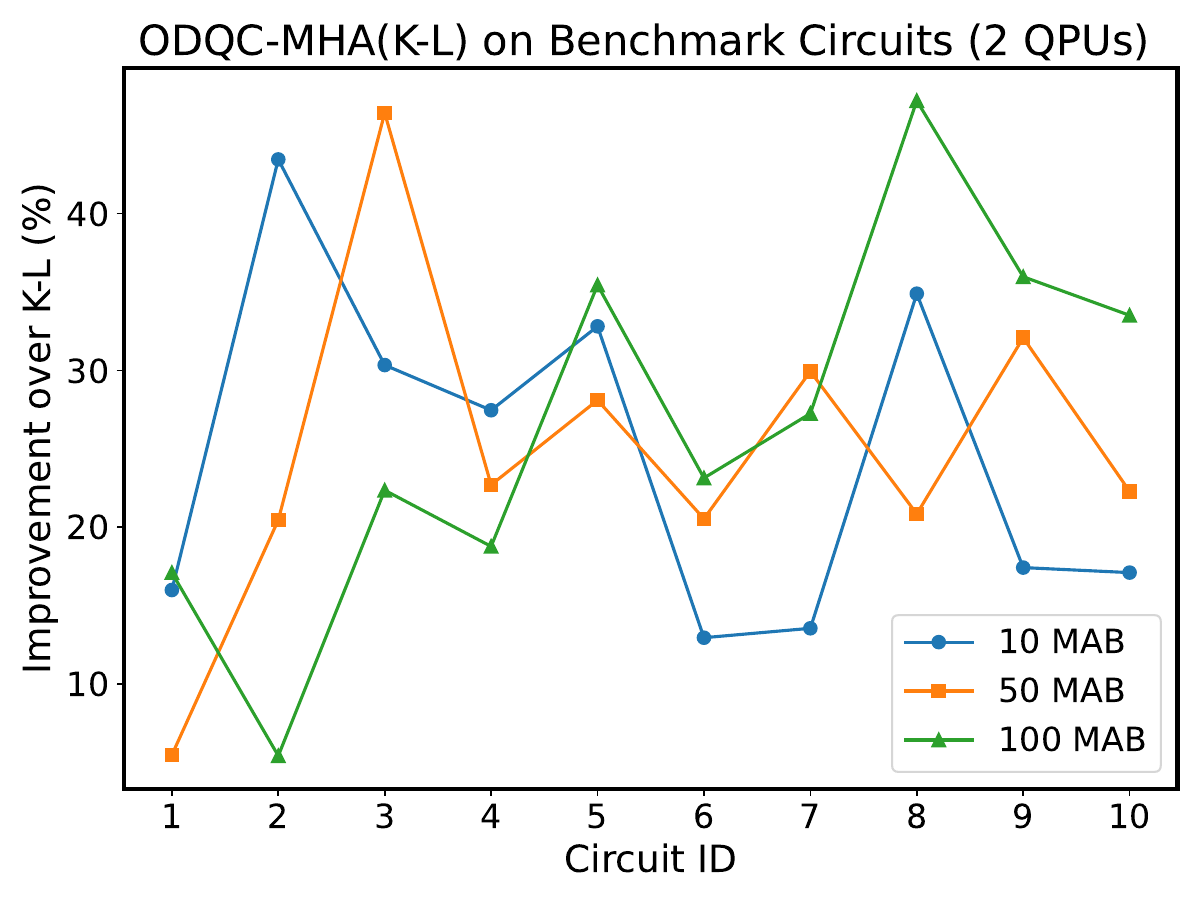}

\caption{Comparison percentage improvement for ODQC-MHA(K-L) on benchmark circuits for varying maximum allowed number of blocks (MAB) (10,50,100), across 2 QPUs. The benchmark circuits are ordered by increasing circuit depth. Each data point is the mean of 100 executions of ODQC-MHA(K-L). Circuit ID \ref{table:circuits}}
\label{fig:non-random}
\end{figure}

Initially, we ran ODQC-MHA(K-L) on quantum circuits from QASMBench \cite{Pnnl}. Information about the benchmark circuits used is shown in Table \ref{table:circuits}. The results of this analysis are shown in Figure \ref{fig:non-random}, showing the percentage improvement over using K-L for the entire circuit i.e. no qubit teleportations. We compare three configurations of ODQC-MHA(K-L), allowing for a maximum allowed number of blocks (MAB) of; 10, 50 and 100. While the amount of improvement varies across the circuits, we see a clear trend. For increasing depth there is a region for which 100 MAB shows the smallest improvement while 10 MAB shows the most, in the middle region we see that 50 MAB shows the most improvement, and for the higher end of circuit depth, 100 MAB. We identify a likely reason for this trend, that for smaller circuits it is possible to 'over-block', that is, it becomes hard to converge on a solution - on average - due to the starting size of an individual genotype. We believe that this analysis on small benchmark circuits is arbitrary because the performance of the heuristic depends strongly on the distribution of CNOT gates, in the next section, we discuss the performance of the algorithm on average, using large randomly generated circuits.

The fact that the performance between different configurations is inconsistent across different circuits suggests a limitation with the design of ODQC-MHA. Future work might include an optimisation of the allowed size of an individual, the population size and number of generations, dependant on the circuit at hand. This should allow the algorithm to search the given solution space more efficiently and prevent converging on local minima. 

\begin{figure}[h!]
\includegraphics[width=0.5\textwidth]{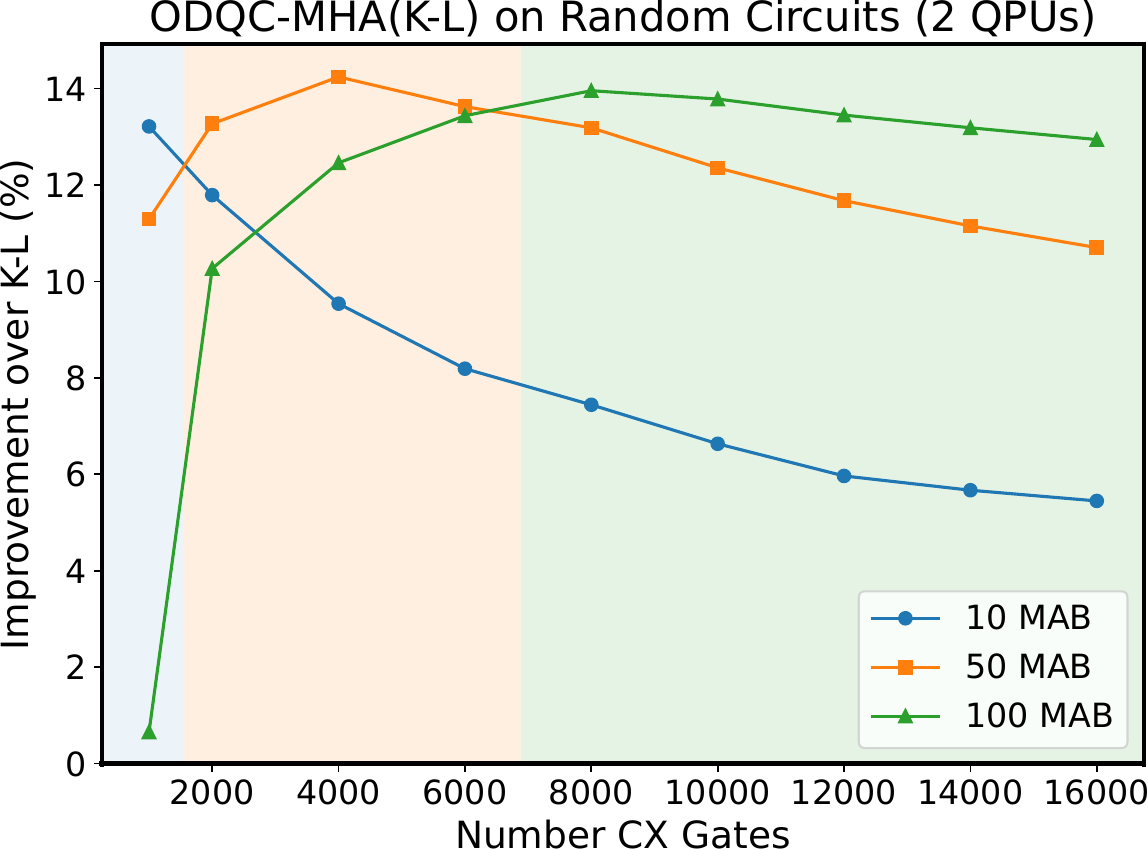}

\caption{Comparison of ODQC-MHA(K-L) for varying maximum allowed number of blocks (10,50,100), to ODQC-MHA(K-L) for 1 block on randomly generated circuits, distributed over 2 processors. Each data point is the mean value of multiple runs of ODQC-MHA(K-L) on different randomly generated, 16 qubit, circuits. The percentage difference is plotted.}
\label{fig:random}
\end{figure}

\begin{figure}[h!]

\centering
\includegraphics[width=0.5\textwidth]{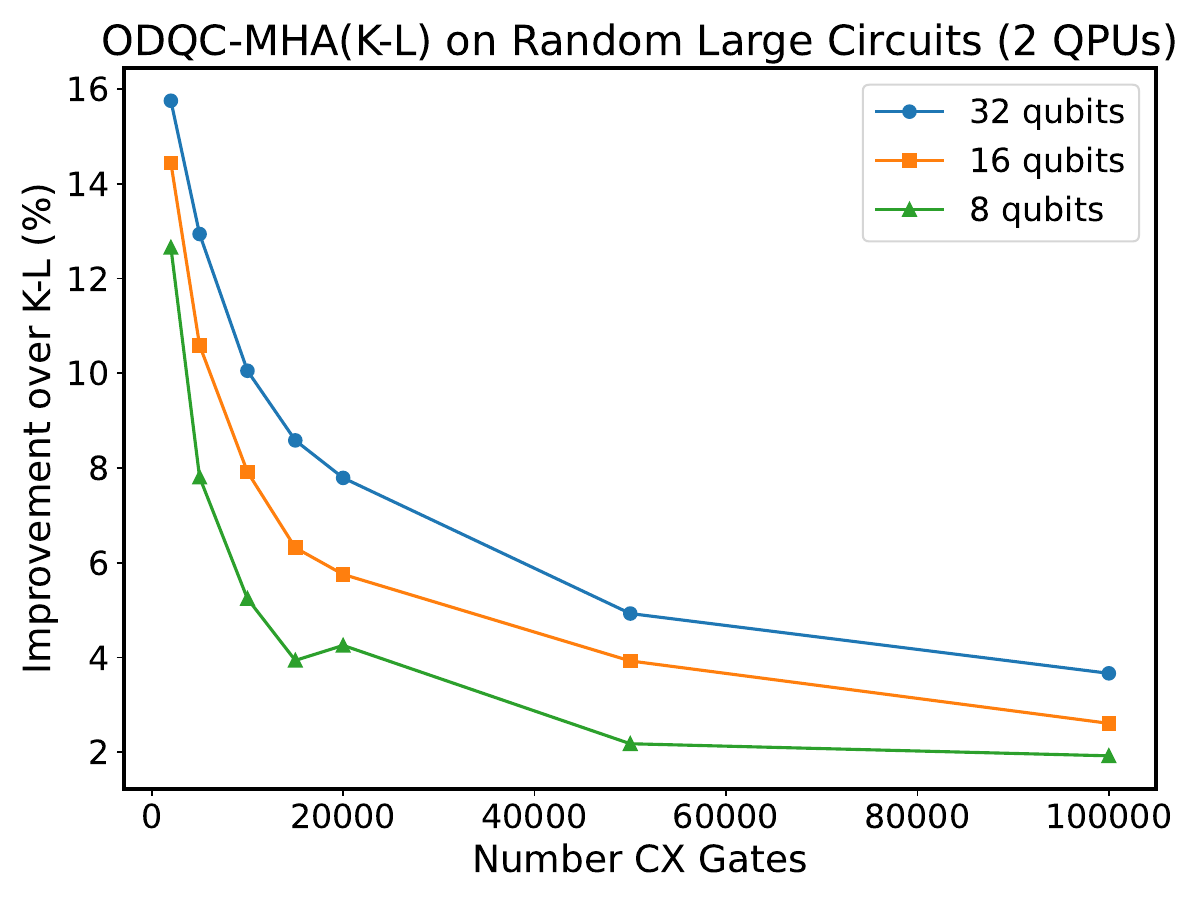}
\caption{Comparison of ODQC-MHA(K-L) to K-L on randomly generated circuits of varying number of qubits (8,16,32) on randomly generated circuits, across 2 QPUs.}
\label{fig:ex3}
\end{figure}

\subsection{Random Quantum Circuits (2 QPUs)}

To demonstrate the effectiveness of ODQC-MHA on average, we ran qubit allocation on randomly generated circuits containing only CNOT gates, which are the only universal gates that are considered in the algorithm. This demonstrates the average behaviour on large circuits where the distribution of CNOT gates tends towards homogeneity. In principal, any bias in the distribution of CNOTs should be exploited by a good heuristic method.

Firstly, we generated random, 16 qubit circuits of varying number of CNOT gates, in the same range as the benchmark circuits used. We ran ODQC-MHA(K-L) to allocate across 2 QPUs for each circuit, this analysis was done 100 times and the average performance is plotted in Figure \ref{fig:random}. This was done for different initial configurations of ODQC-MHA(K-L) allowing for a maximum number of blocks (length of genotpye) of 10,50 and 100. It is clear that the genetic algorithm was able to - on average - converge on a solution with fewer bell pairs required across all the randomly generated circuits, with a general trend towards smaller improvement for increasing number of gates. However, we again observe the same trend as explained before, a region where each configuration performs best. This effect is shown dramatically in the first points (left). This indicates that ODQC-MHA(K-L) requires further optimisation to select an optimal number of blocks. Due to resource limitations, we were not able to analyse the limit that an improvement is shown by increasing the allowed maximum number of blocks.

Next, we generated random circuits of; 8, 16, and 32 qubits with up to 100,000 CX gates, in order to test the performance of ODQC-MHA(K-L) on larger circuits.The results are plotted in Figure \ref{fig:ex3}. We see a similar improvement trend across different number of gates. Interestingly, the reduction in the number of bell pairs increases for greater number of qubits. This could be because for the same number of CNOT gates across more qubits, the interactions are more sparsely distributed and will likely have shorter depth.

\subsection{Random Quantum Circuits (3 QPUs)}

Finally, to demonstrate the performance of ODQC-MHA using a different heuristic for graph partitioning within each block, we performed the same analysis but distributing each circuit across 3 QPUs using ODQC-MHA(GPA) \ref{greedy} within each block. Here we observe a similar trend as across 2 QPUs, however, the first point for maximum allowed blocks of 100 shows anomalously high improvement. We can attribute this to the smallest circuits having larger variance in distribution of CX gates and so the genetic algorithm may get 'lucky' on certain circuit configurations. 
\begin{figure}
\includegraphics[width=0.5\textwidth]{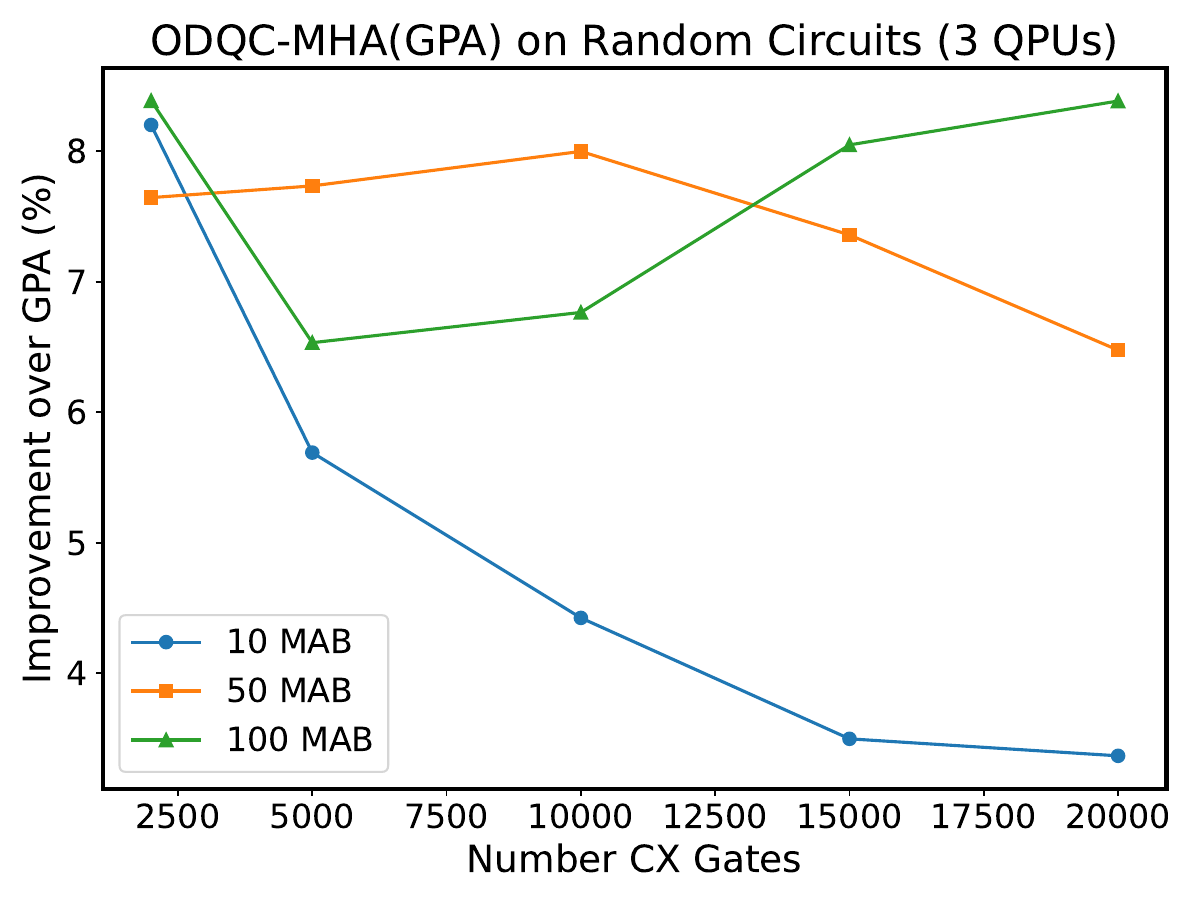}
\caption{Comparison of ODQC-MHA(GPA) to GPA on randomly generated 16 qubit circuits, for different maximum allowed number of blocks (10,50,100)}
\label{fig:ex4}
\end{figure}

\subsection{Discussion}
In the limit that the number of gates is large, we would expect that the performance of the K-L algorithm at allocating the qubits across two processors would tend to the performance of randomly allocating the qubits across two processor. This would mean that half of the CNOT gates are executed remotely. However, for any finite circuit depth, the optimal solution will always be better than half of the CNOT gates. This also applies to our meta heuristic, because the genetic algorithm can always find a block length where the performance of K-L is better than half and so - if allowed to execute properly - will be able to exploit sections of the circuit where K-L performs well at allocating the qubits. 




\section{Conclusions}
\label{sec:conclusion}

Compiling circuits for DQC will be paramount to the future and scaling of quantum computers, within a 'quantum data center'. In this letter we addressed the problem of qubit allocation within a compilation, using a meta-heuristic which optimises for the minimum number of qubit and gate teleportations. The results when comparing our method to the standard method of graph partitioning using the K-L algorithm shows a significant improvement.

\section{Acknowledgements}

The research work was supported by the Army Research Office MURI under the project number W911NF2110325, by the QuantumCT project, and by the National Science Foundation under project number CNS 2402862.


\bibliographystyle{IEEEtran}
\bibliography{references3}
\end{document}